\documentclass[useAMS,usenatbib]{mn2e}
\usepackage{aas_macros}
\usepackage[dvips]{graphicx}
\usepackage{amsmath,amssymb}
\title[Cusp-core transformations induced by AGN feedback]{Cusp-core transformations  induced by AGN feedback in the progenitors of cluster galaxies}
\author[D. Martizzi et al.]{\parbox[t]{\textwidth}{Davide Martizzi$^{1}$\thanks{E-mail: martdav@physik.uzh.ch}, 
Romain Teyssier$^{1,2}$ and Ben Moore$^{1}$}\vspace*{6pt}\\
$^{1}$Institute for Theoretical Physics, University of Zurich, CH-8057 Z\"urich, Switzerland\\
$^{2}$CEA Saclay, DSM/IRFU/SAP, B\^atiment 709, F-91191 Gif-sur-Yvette, Cedex, France}

\begin{document}

\maketitle

\label{firstpage}

\begin{abstract}
In a recent study (Martizzi et al. 2012), we used cosmological simulations to show that active galactic nuclei (AGN) feedback on the gas distribution in clusters of galaxies can be important in determining the spatial distribution of stars and dark matter in the central regions of these systems. The hierarchical assembly of dark matter, baryons and black holes obscures the physical mechanism behind the restructuring process. Here we use idealized simulations to follow the response of a massive dark matter halo as we feed the central black hole with a controlled supply of cold gas. This removes most of the complexity taking place in the cosmological simulations that may have biased our previous study. We confirm our previous results: gas heated and expelled from the central regions of the halo by AGN feedback can return after cooling; repeated cycles generate gravitational potential fluctuations responsible for irreversible modifications of the dark matter mass profile. The main result is the expulsion of large amounts of baryons and dark matter from the central regions of the halo. {According to the work presented here, outflow induced fluctuations represent the only mechanism able to efficiently create dark matter cores in clusters of galaxies.}
\end{abstract}


\section{Introduction}

In the concordance cosmological scenario, the large scale structure of the universe is globally determined by the dominant mass component, dark matter. The situation is different within highly non-linear regions where the baryons dominate. Gas cooling, star formation, feedback processes and dynamical interactions between different mass components can modify the spatial distributions of all components. For example, dark matter halos are known to globally respond to the condensation of baryonic matter in their central regions via dissipative processes. This increases the central mass density through adiabatic contraction, and causes the halo to become rounder. More recent studies of baryonic processes have focussed on the effects of supernovae feedback in galaxies \citep{Governato:2010p1442, 2011arXiv1111.5620M, 2012arXiv1206.4895T}, however there are fewer studies of the effects of AGN feedback on massive galaxies.

Clusters of galaxies provides a wealth of opportunities to test the validity of the current paradigm for cosmic structure formation. Analysing the details of the spatial distribution of the different components of a cluster, as predicted by theory, and comparing them to those measured (or inferred) from observations allows us to test our current understanding of high mass galaxy formation. 

In this paper we focus on the role of active galactic nucleus (AGN) feedback in shaping the mass distribution of clusters of galaxies. The main motivation for this work comes from recent observational constraints of cluster mass profiles using gravitational lensing: \cite{2004ApJ...604...88S}, \cite{2008ApJ...674..711S}, \cite{2009ApJ...706.1078N}, \cite{2011ApJ...728L..39N}, \cite{2011A&A...531A.119R}, \cite{2012arXiv1209.1391N}. These report mass density profiles whose central slopes, measured within the inner $\approx$ 5 kpc, are much shallower than the universally adopted NFW profile. On a slightly smaller scale, within 1-2 kpc from the centre, very shallow slopes in the stellar surface brightness profiles of massive elliptical and cD galaxies have been observed \citep{1999ASPC..182..124K, 2000ApJS..128...85Q, 2003AJ....125..478L, 2004AJ....127.1917T, 2005AJ....129.2138L, 2007ApJ...671.1456C,  2009ApJS..182..216K, 2011arXiv1108.0997G}. The combination of these two observational facts suggests a possible connection between the mechanisms that produce shallow density slopes in the dark matter distribution and in the stellar distribution. For example, both the dark matter and stellar components can respond similarly to dynamical perturbations since they behave like collisionless fluids. 

{The formation of cores with a shallow central density profile is currently a challenge for the concordance cosmological scenario. Dissipationless collapse and hierarchical merging always gives rise to cuspy central density profiles. Alternative dark matter candidates, including self interactive dark matter, seem to provide a viable mechanism to form dark matter cores \citep{2012arXiv1208.3025R, 2012MNRAS.423.3740V}. In addition, \cite{2012MNRAS.424..747L} showed that dynamical heating obtained through a long series of dry mergers can produce density cores in galaxy clusters, even in the context of the standard cosmological scenario. Similar results are discussed in \cite{2009ApJ...698.2093D} and \cite{2012MNRAS.424...38D}. However, there is growing theoretical consensus that processes related to baryonic physics may play a dominant role in the formation of dark matter cores  \citep{2006Natur.442..539M, 2008A&A...479..123P, 2008Sci...319..174M, Governato:2010p1442, 2010MNRAS.405.2161D, 2011arXiv1111.5620M, 2012MNRAS.421.3464P, 2012arXiv1206.4895T, 2012arXiv1202.1527R}. Of particular interest is \cite{2012MNRAS.421.3464P}, who use an anlytical model to study the dynamical effects of potential fluctuations. According to this scenario, in low mass halos the energy injected by supernovae explosions is sufficient to erase dark matter cusps via the generation of repeated impulsive gas outflows responsible for rapid potential fluctuations \citep{2012ApJ...759L..42P}. In galaxy clusters, supernovae feedback does not provide enough energy to unbind the gas, however AGN feedback is expected to be strong enough to generate repeated gas outflows in their central regions.} 

In a recent study \citep{2012MNRAS.422.3081M}, we used high resolution cosmological hydrodynamical simulations including AGN feedback to show that it is possible to produce clusters with similar properties as indicated by these observations. We argued that at least three processes could contribute to creating shallow density profiles: (i) supermassive black holes (SMBHs) can transfer part of their orbital energy to collisionless matter via dynamical friction during mergers; part of the collisionless matter is expelled from the central regions (i.e. \cite{2010ApJ...725.1707G}); (ii) AGN driven gas outflows modify the gravitational potential close to SMBHs; ejection of collisionless matter from the central region of the cluster is produced if the potential fluctuations happen on a time-scale smaller than, or resonant with, the dynamical time. To be effective this mechanism requires many subsequent potential fluctuations (\cite{1996MNRAS.283L..72N, 2002MNRAS.333..299G, 2005MNRAS.356..107R, 2012MNRAS.421.3464P}); (iii) slow loss of gas mass from the central regions of a cluster can produce an expansion of the inner mass distribution; this slow loss of gas can be obtained as an effect of AGN heating. 

Our previous results indicated that process (ii) was the dominant contribution to the shallowing of our central dark matter and stellar cusps. Very similar results were also found by \cite{2010MNRAS.405.2161D} using the OWLS suite of cosmological simulations. Related mechanisms have been found to be active in lower mass halos \citep{Governato:2010p1442, 2012MNRAS.421.3464P, 2011arXiv1111.5620M, 2012arXiv1206.4895T}. Unfortunately, the complexity of hierarchical structure formation makes it difficult to disentangle the efficiency of this mechanism. To obtain a clearer understanding we will use simpler, idealised simulations, that remove these ambiguities in order to provide a better insight into the core creation process.

In a recent paper, \cite{2012arXiv1202.1527R} study the role of baryonic mass outflows in galaxy clusters, using collisionless matter simulations where gas is modeled as a time varying external potential. In this paper we choose a similar approach, but we choose to perform full N-body+hydrodynamical simulations of a isolated dark matter halo, including gas cooling and AGN feedback from a single central supermassive black hole. AGN heating can produce the impulsive expulsion of a large quantity of gas from the central regions of a cluster. The result of this is a fast varying central gravitational potential without the dynamical effects of hierarchical mergers of galaxies and their black holes.
 
The paper is organized as follows: the first section is dedicated to the numerical methods we adopted for our simulations; the second section gives our the main results; in the last section we discuss and provide a short summary.

\section{The simulations}
\label{sec:thesims}

We use the AMR code RAMSES \citep{Teyssier:2002p451} to perform simulations that follow the evolution of the mass distribution of an isolated dark matter halo in presence of a gaseous component. The simulations differ in the way initial conditions for the gas are set, as explained in subsection \ref{subsec:ics}. The gas dynamics is modeled using a second-order unsplit Godunov scheme \citep{Teyssier:2002p451, Teyssier:2006p413, Fromang:2006p400} based on the HLLC Riemann solver and the MinMod slope limiter \citep{Toro:1994p1151}. We assume a perfect gas equation of state (EOS) with $\gamma=5/3$. We sample the dark matter distribution with $10^6$ particles of mass $m_{\rm dark}=1.78 \times 10^7$ M$_{\odot}$. In all simulations gas represents 17\% of the total mass in the halo; the gas mass resolution element is $m_{\rm gas}=3.79 \times 10^6$ M$_{\odot}$. The minimum cell size is $\Delta x_{\rm min} = 0.18$ kpc. The AMR grid was dynamically refined using a quasi-Lagrangian strategy: when the dark matter or baryons mass in a cell reaches 8 times the mass resolution, it is split into 8 children cells. 

Gas is allowed to cool: we use the \citealt{sutherland_dopita93} cooling function to account for H, He and metal cooling. We include AGN feedback, using the same implementation we adopted in \cite{2011MNRAS.414..195T}, \cite{2012MNRAS.420.2859M} and \cite{2012MNRAS.422.3081M}, a modified version of the \cite{Booth:2009p501} model. SMBHs are modeled as sink particles, following \cite{Krumholz:2004p1079}. The gas accretion onto SMBHs is computed using a modified Bondi-Hoyle formula \citep{Booth:2009p501, 2012MNRAS.420.2662D}. A fraction of the accreted mass is converted into thermal energy that is directly injected into the gas surrounding the black hole. Despite its simplicity, when properly tuned, the model reproduces basic features of AGN feedback (as it is thought to work), in particular the existence of two feedback modes: the impulsive 'quasar mode' during cold gas accretion when the accretion rate is high, and the quiescent 'radio mode' during hot gas accretion when the accretion rate is low. In this picture, the effect of 'radio mode' feedback is to prevent hot gas from cooling, while the effect of 'quasar mode' feedback is to eject gas from galaxies, violently suppressing star formation.

\subsection{Initial conditions}
\label{subsec:ics}

The initial conditions for dark matter are scaled-up versions of those adopted by \cite{2012arXiv1206.4895T}. The dark matter distribution follows the NFW density profile. We chose a concentration parameter $c=10$ and a circular velocity $V_{200}=350$ km/s, corresponding to a virial mass $M_{200}=1.42 \times 10^{13}$ M$_{\odot}$ and a virial radius $R_{200}=500$~kpc. We truncate the halo at 1.13 Mpc, so that the total enclosed mass is $M_{\rm halo}=1.78 \times 10^{13}$ M$_{\odot}$. Progenitors of halos of mass $\sim 10^{14}$ M$_{\odot}$ identified at redshift $z=0$ have a typical mass of a few $10^{13}$ M$_{\odot}$ at redshift $z\gtrsim 2$. This means that our simulations are particularly suited for comparing with the early evolution of clusters of galaxies. We choose the initial velocity field of the particles in a way that equilibrium can be maintained and such that the system has spin parameter $\lambda=0$ (no rotation). This choice has been made to avoid dealing with the details of gas accretion onto SMBHs in a clumpy disk \citep{2012ApJ...757...81B} and it allows us to study test cases in which a high gas accretion efficiency is reached. 

In all simulations a gaseous halo is embedded in the dark matter halo. The gas density $\rho$, pressure $P$ and temperature $T$ follow specific profiles:
\begin{eqnarray}
 \rho(x)&=&\rho_0\left[\frac{\ln(1+x)}{x}\right]^{\frac{1}{\Gamma-1}} \\
 P(x)&=&4\pi G\rho_0\rho_s r_s^2 \frac{\Gamma-1}{\Gamma}\left[\frac{\ln(1+x)}{x}\right]^{\frac{\Gamma}{\Gamma-1}} \\
 T(x)&=&T_0\frac{\ln (1+x)}{x} 
\end{eqnarray}
where $r_s$ is the NFW scale radius, $\rho_s$ is the dark matter density at $r_s$, and $x=r/r_s$. We set the central temperature $T_0$ according to 
\begin{equation}
\frac{k_B T_0}{\mu m_p}=4\pi G\rho_s r_s^2 \frac{\Gamma-1}{\Gamma}.
\end{equation}
Note that our choice implies that $P\propto \rho^{\Gamma}$. The effective polytropic index $\Gamma$ is set according to 
\begin{equation}
\Gamma=1+\frac{(1+x_{eq})\ln(1+x_{eq})-x_{eq}}{(1+3x_{eq})\ln(1+x_{eq})},
\end{equation}
where $x_{eq}=\sqrt{5}c$. See \cite{2001MNRAS.327.1353K} for a similar, but slightly more complex model. With our choice of parameters, the gas density profile follows the slope of the NFW profile for $x\geq x_{eq}$ and tends to a constant value $\rho_0$ for $x\ll x_{eq}$.The central density $\rho_0$ is set requiring the gaseous halo mass fraction to be $f_{\rm gas}=0.15$. The metallicity of the gaseous halo is set to $Z=10^{-2}Z_{\odot}$. The stability of the initial conditions was confirmed by running adiabatic simulations (no gas cooling).  

Within the halo we place ten spherical gas clumps of equal mass, $M_{\rm cl}=2.1\times 10^{10}$ M$_{\odot}$ and radius $R_{\rm cl}=10$ kpc. We set the temperature of the gas in the clumps to be a few $10^4$ K; its metallicity is set to $Z=Z_{\odot}$. Each clump is given an initial velocity of magnitude $V_{\rm cl}=V_{200}$ directed towards the halo center. The aim of these clumps is to roughly mimic the cold gas reservoirs provided by infalling galaxies into a cluster. We have chosen radial orbits to be able to efficiently provide gas to the central regions of the halo; such a choice allows us to study the maximum effect gas dynamics can have on the dark matter distribution. 

Our simulations differ in the initial geometrical distribution of the clumps. In one simulation, the distance of the clouds from the centre is in the range $200 \hbox{ kpc} \lesssim r \lesssim 1 \hbox{ Mpc}$; in this case the clouds far from the centre and reach it slowly, we labeled this simulation as SLOW. In the second simulation the distance of the clouds from the center is in the range $50 \hbox{ kpc} \lesssim r \lesssim 250 \hbox{ kpc}$; since the clouds are close to the centre and reach it faster, we labeled this simulation as FAST. We also consider a third run in which the clumps distribution is the same as the SLOW case, but no AGN feedback is considered; this run has been labeled as AGNOFF. 

Finally, as additional test cases we run two simulations which only include the dark matter and smooth gaseous halos. The first one includes AGN feedback and is labeled as SMOOTH AGNON, the second one does not include AGN feedback and is labeled as SMOOTH NOAGN. In these two runs the halo metallicity is set to a higher value $Z=0.2 Z_{\odot}$. These latter two simulations allow us to study the effects of a cooling flow, rather than clumpy accretion.

In the FAST, SLOW and SMOOTH AGNON simulations, we place a SMBH of mass $M_{\rm BH}=10^9$ M$_{\odot}$ at the centre of the halo. This choice for the black hole mass roughly matches the value measured at redshift $z\approx2$ for the central SMBH in the cosmological simulation analysed in \cite{2012MNRAS.422.3081M} and agrees with the expected mass from the $M_{\rm BH}-\sigma$ relation \citep{tremaine_etal04}.

\section{Results}
\label{sec:results}
\begin{figure*}
    \includegraphics[width=0.99\textwidth]{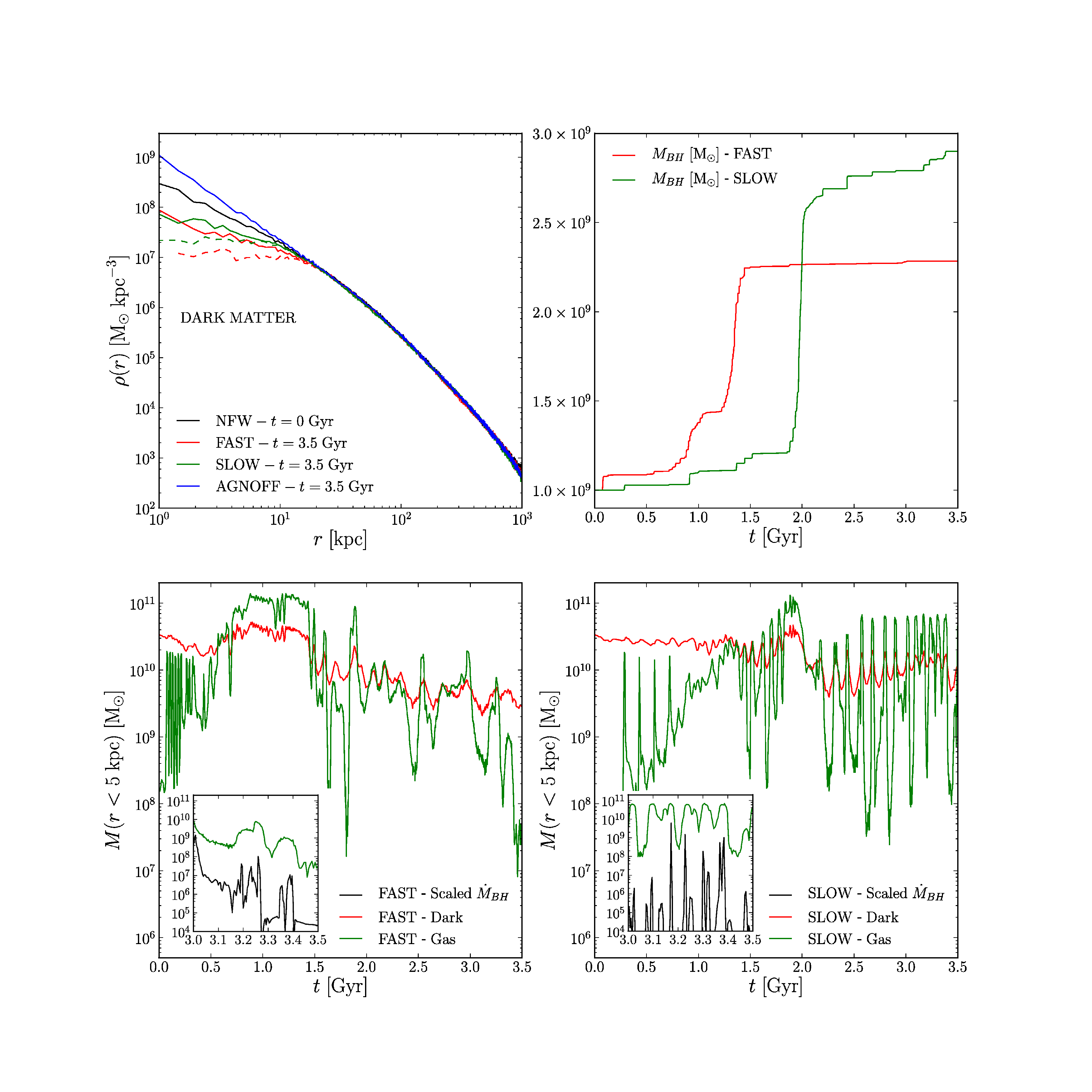}
  \caption{Top left: dark matter density profiles at different times and for different simulations. After 3.5 Gyr a large amount of mass has been removed from the central region in the FAST and SLOW simulations. The AGNOFF simulation shows signs of adiabatic contraction, with a large accumulation of mass in the central region. The solid lines are obtained when the peak of the dark matter distribution is chosen as a center, whilst the dashed lines are obtained when the SMBH is chosen as the center. Top right: SMBH mass evolution in the FAST and SLOW simulations. The mass evolves differently in the two runs, but the qualitative evolution through bursts of Eddington limited gas accretion is the same. Bottom panels: evolution of the dark (red line) and gaseous (green line) mass within 5 kpc from the cluster center in the FAST (left) and SLOW (right) simulations. After 2 Gyr a large amount of dark matter is removed from the central region. The inset panels show a zoomed view of the evolution of the gas mass within 5 kpc, which correlates with the evolution of the accretion rate. As soon as gas is pushed away from the SMBH by AGN feedback the accretion rate drops.}
  \label{fig:everything}
\end{figure*}


We run our simulations for 3.5 Gyr. For comparison, the dynamical time in the central region of the halo ($r<5$ kpc), which depends on the density and radius,  is of the order of $10^8$ yr. As shown by the top left panel of Figure \ref{fig:everything}, 3.5 Gyr is sufficient time to see convergence in the effects of dark matter removal from the central regions of the halo. The black continuous line shows the initial mass profile in both simulations. After 1--2 Gyr a significant amount of mass has been ejected from the region $r<5$ kpc in both the FAST (red line) and SLOW (green line) simulations. 

The convergence to a central inner core occurs in both simulations with AGN feedback. On the contrary, the AGNOFF simulation (blue line) is completely different: as the clumps fall to the central region of the cluster and as the gaseous halo cools, the dark matter halo contracts as a response to gas condensation and as a result the central density increases. 

Some differences can be observed when considering the evolution of the mass of the central SMBH as a function of time $t$ in our models (top right panel of Figure \ref{fig:everything}). As expected, in the FAST simulation the SMBH starts growing much earlier. The final black hole masses are $M_{BH}= 2.28\times 10^{9}$ M$_{\odot}$ and $M_{BH}=2.86\times 10^{9}$ M$_{\odot}$ for the FAST and SLOW simulations respectively. In the FAST simulation the gas clumps feed the SMBH earlier, triggering a phase of rapid mass growth ($t<1.5$ Gyr) followed by a phase of intense AGN feedback. This is then responsible for the subsequent slowing down of mass accretion onto the SMBH. In the SLOW simulation, the gas clumps start feeding the SMBH later, so the early evolution of its mass is mild ($t<1.8$ Gyr); at later times a large quantity of gas becomes available to accrete onto the SMBH, triggering a very rapid mass growth in the time interval $1.8$~Gyr~$<t<2.1$~Gyr; this phase is also followed by intense AGN activity which slows down the SMBH mass growth.

The bottom panels of Figure \ref{fig:everything} show the evolution of the gas and dark mass enclosed within 5 kpc from the halo centre as a function of time. This is useful for understanding the effect of AGN feeback in the central region. The passages of the gas clumps close to the centre during the early evolution can be detected as peaks in the enclosed gas mass ($t<0.7$ Gyr for FAST and $t<1.4$ Gyr for SLOW). Before the central SMBH becomes massive enough to unbind the gas provided by the clumps, a large quantity of mass is accumulated in the central region. This produces a contraction of the total mass distribution that can be observed in this plot as a steep increase in the gas and dark mass within 5 kpc. 

The striking effect we observe is that this contraction is completely erased as soon as the AGN activity becomes violent enough to unbind the gas. After the first passage of the clumps through the centre, the high metallicity material mixes with the low metallicity gas in the halo. By this time, a large reservoir of cooling gas is available. This gas is feeding the SMBH but is also influenced by its feedback. AGN heating results in the ejection of gas from this region. The ejected gas eventually cools down and begins infalling again. The accretion-ejection cycle happens on a timescale that is always comparable to the dynamical time in the considered region. Fluctuations of the dark matter mass enclosed in this region follow the fluctuations in the gas mass. The result of subsequent accretion-ejection cycles is the irreversible ejection of $\sim 2 \times 10^{10}$ M$_{\odot}$ from $r<5$ kpc, which is roughly 60~\% of the dark mass initially enclosed in this region. 

There are interesting differences in the properties of the accretion-ejection cycles in the two simulations: they are longer and more irregular in the FAST case, but they appear to be quasi-periodic in the SLOW case. This difference is possibly due to the fact that for $t>2$ Gyr the SMBH in the SLOW simulation can provide a more efficient feedback due to its larger mass. 

\begin{figure}
    \includegraphics[width=0.50\textwidth]{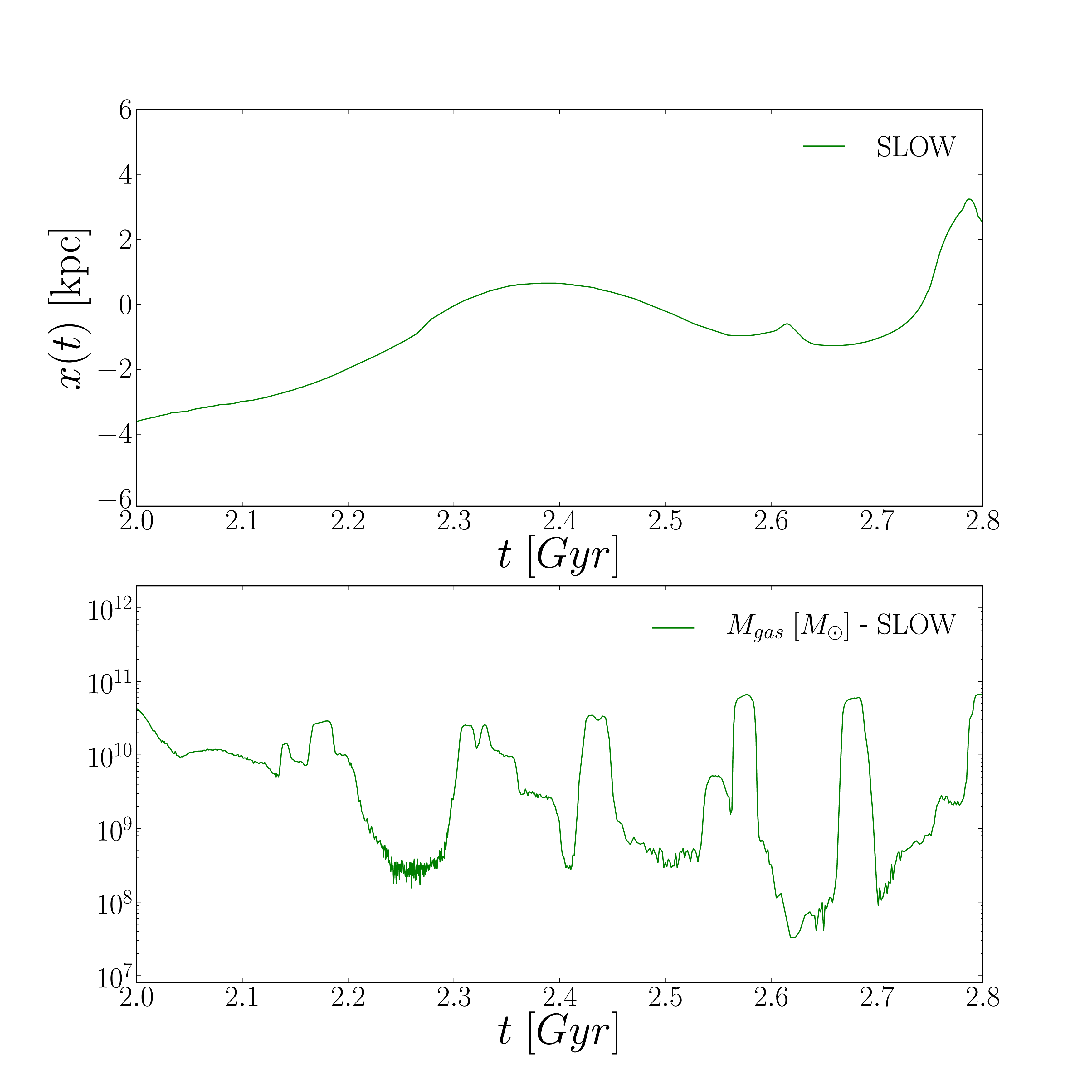}
  \caption{Top panel: distance of the SMBH from the centre of the dark matter distribution as a function of time. Bottom: gas mass within 5 kpc as a function of time. The SMBH is chosen as the centre. This plot shows that the decrease in the gas mass poorly correlates with the black hole motions. Therefore, the decrease in gas mass is not a consequence of the SMBH moving to lower density regions but rather due to AGN feedback.}
  \label{fig:trajectory}
\end{figure}

The inset panels in Figure \ref{fig:everything} show the correlation between the accretion rate onto the SMBH and the amount of gas in the centre. Whenever gas mass in the centre is removed, the accretion rate suddenly decreases. To check whether this is a real effect or just a product of the black hole oscillating within a perturbed potential or within a harmonic core, we compare the distance of the SMBH from the dark matter centre as a function of time to the evolution of the gas mass within 5 kpc during the same time interval (Figure \ref{fig:trajectory}). This plot is made using the SLOW simulation and a similar result is found in the FAST case. There is no obvious correlation between these quantities, implying that the adopted centering technique is not biasing our results. {Furthermore, the mass of the dark matter particles is $\gtrsim 50-100$ times smaller than the mass of the SMBH and for most of the time gas dominates the potential in the central region of the halo. We can therefore exclude spurious two body interactions from driving the black hole motions. }

\begin{figure*}
    \includegraphics[width=0.99\textwidth]{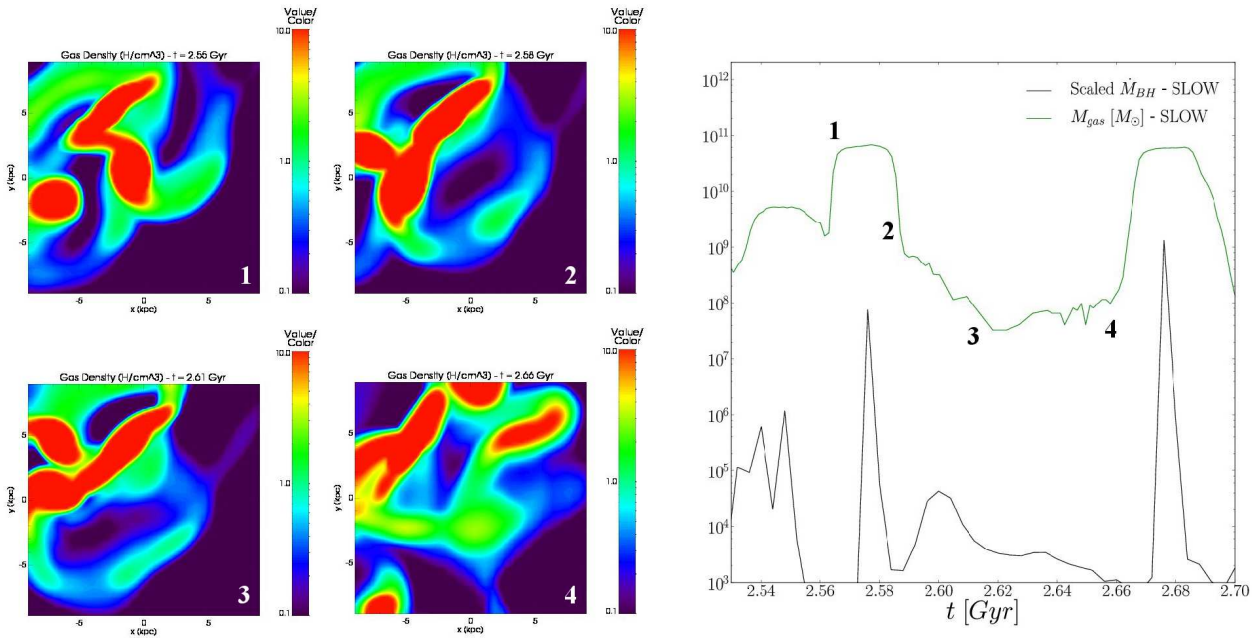}
  \caption{Projected density maps are shown for four snapshots over an accretion event followed by strong feedback. The SMBH is located at the centre of each map. This time sequence illustrates how AGN activity is able to excavate a hole in the gas density that is replenished on a time scale of $\sim 10^8$ Gyr. The right panel shows the evolution of the gas mass within 5 kpc from the SMBH together with the gas accretion rate (rescaled).}
  \label{fig:maps}
\end{figure*}

Figure \ref{fig:maps} shows a time sequence of images of the gas distribution over an interval of 100 million years. This spans the period in which the SMBH dramatically lowers the central gas density. The left panels show four projected gas density maps at four different times, before, during and after an intense AGN activity burst. The SMBH is located at the centre of each map. The right panel of the same figure shows the evolution of the gas mass within 5 kpc (green) from the SMBH and the re-scaled gas accretion rate (black). The times of the above four snapshots in the maps are labeled as numbers in the right panel. We can clearly see that a large amount of gas is available for accretion in snapshot 1 and this is observed in the right-hand plot as large values of the accretion rate and enclosed gas mass within 5 kpc. In snapshots 2 and 3, AGN feedback blows gas away from the central region and the accretion rate decreases. Finally, in snapshot 4 we see gas returning to the centre and beginning to accrete again onto the SMBH.

\begin{figure}
    \includegraphics[width=0.50\textwidth]{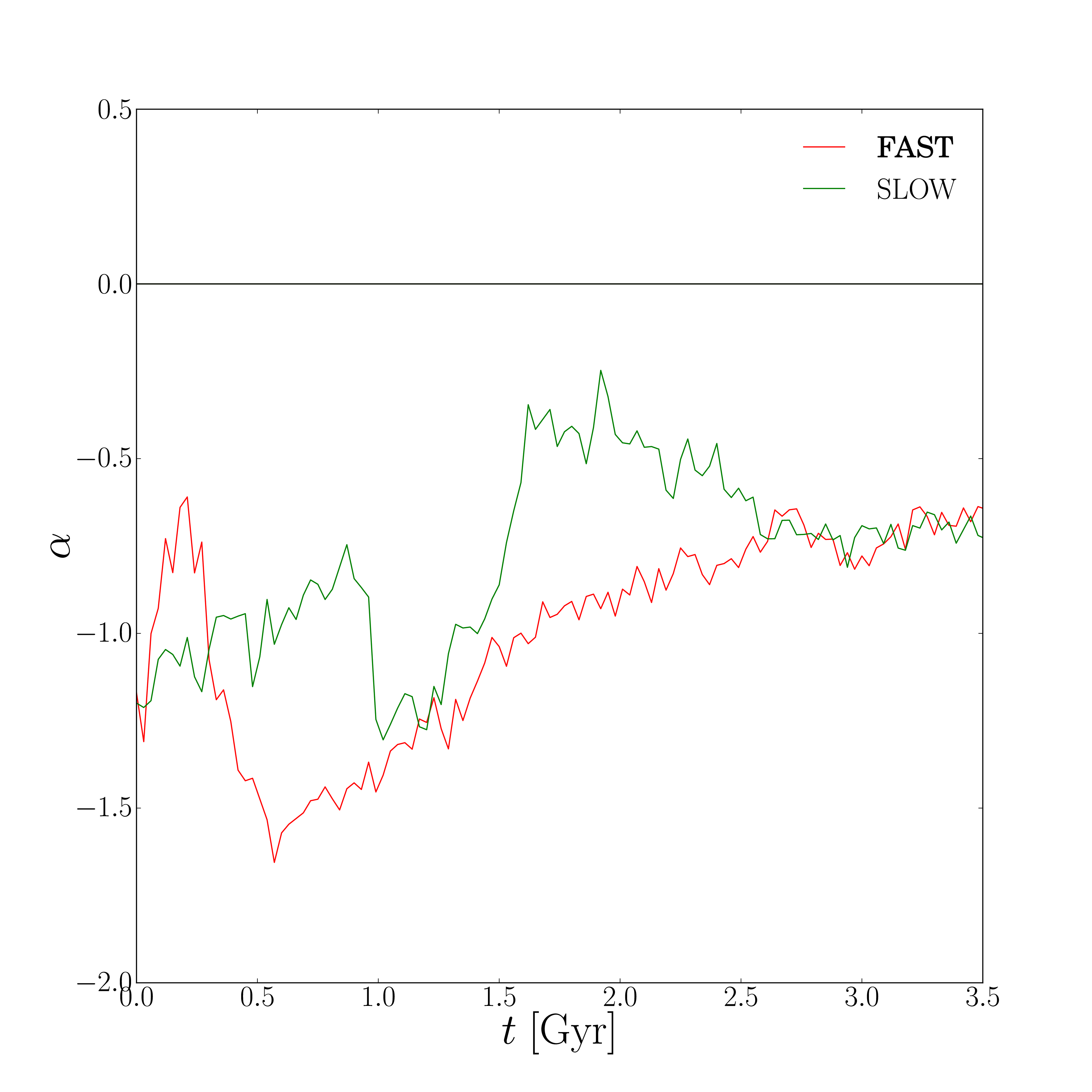}
    \caption{Evolution of the central slope of the dark matter profile as a function of time when the dark matter density peak is chosen as a centre.} 
  \label{fig:slopes}
\end{figure}

In Figure \ref{fig:slopes} we plot $\alpha=d\log(\rho)/d\log(r)$, the central slope of the dark matter density profiles as a function of time. We fit a power law to the density profiles in the region $0.4$~kpc~$<r<8$~kpc. The dark matter peak location has been measured using a shrinking spheres algorithm. A slope of $\alpha=-0.7$ is found if we choose the dark matter peak as the centre. {As it is well known, inferred density profiles are sensitive to the choice of centre: assuming that the black hole is located exactly at the centre of the cluster may lead to slightly different estimates of the central slope. This fact has to be considered as an important caveat for observationally inferred density profiles. }
 
Both the FAST and SLOW simulations converge to the same final density profiles and their evolution over time is qualitatively similar. Initially, the clumps transfer their energy to dark matter via dynamical friction; the result of this energy transfer is an initial shallowing of the slope of the dark matter profile. This process has been shown to be efficient by e.g. \cite{2004ApJ...607L..75E}. {Recently, \cite{2012MNRAS.424..747L} showed that dynamical heating caused by dry mergers in a cosmological context can lead to a similar effect.} However, in our simulations as gas in the halo cools, this adiabatically contracts the dark matter and steepens the central profile, reversing the initial flattening. This period continues until the black hole has grown massive enough to halt the accretion process. The steepest slopes are obtained at $t\sim 0.6$ Gyr and $t\sim 1.2$ Gyr in the FAST and SLOW simulations respectively, as expected from the difference between the clump arrival times in the two runs. 

As soon as AGN feedback becomes strong enough to unbind large quantities of gas, the accretion-ejection cycles start to be efficient in expelling dark matter from the central region with the final result of producing shallower density slopes. For $0.6$~Gyr~$<t< 3.5$~Gyr the value of $\alpha$ in the FAST simulation increases with time. On the contrary, in the SLOW case, $\alpha$ only increases in the interval $1.2$~Gyr~$<t<1.7$~Gyr, and then remains approximately stable for $t>1.7$~Gyr. This difference is caused by the different SMBH mass evolution observed in the two runs which makes AGN feedback operate in different modes, as already observed in Figure \ref{fig:everything}. Despite these differences, the final state of the system is very similar in both the FAST and SLOW cases, demonstrating that once AGN feedback is properly triggered and repeated accretion-ejection cycles are produced, the final configuration of the system is weakly dependent on the initial conditions.

\subsection{Experiments with smooth gas distributions}
\label{sec:smooth}
\begin{figure*}
    \includegraphics[width=0.99\textwidth]{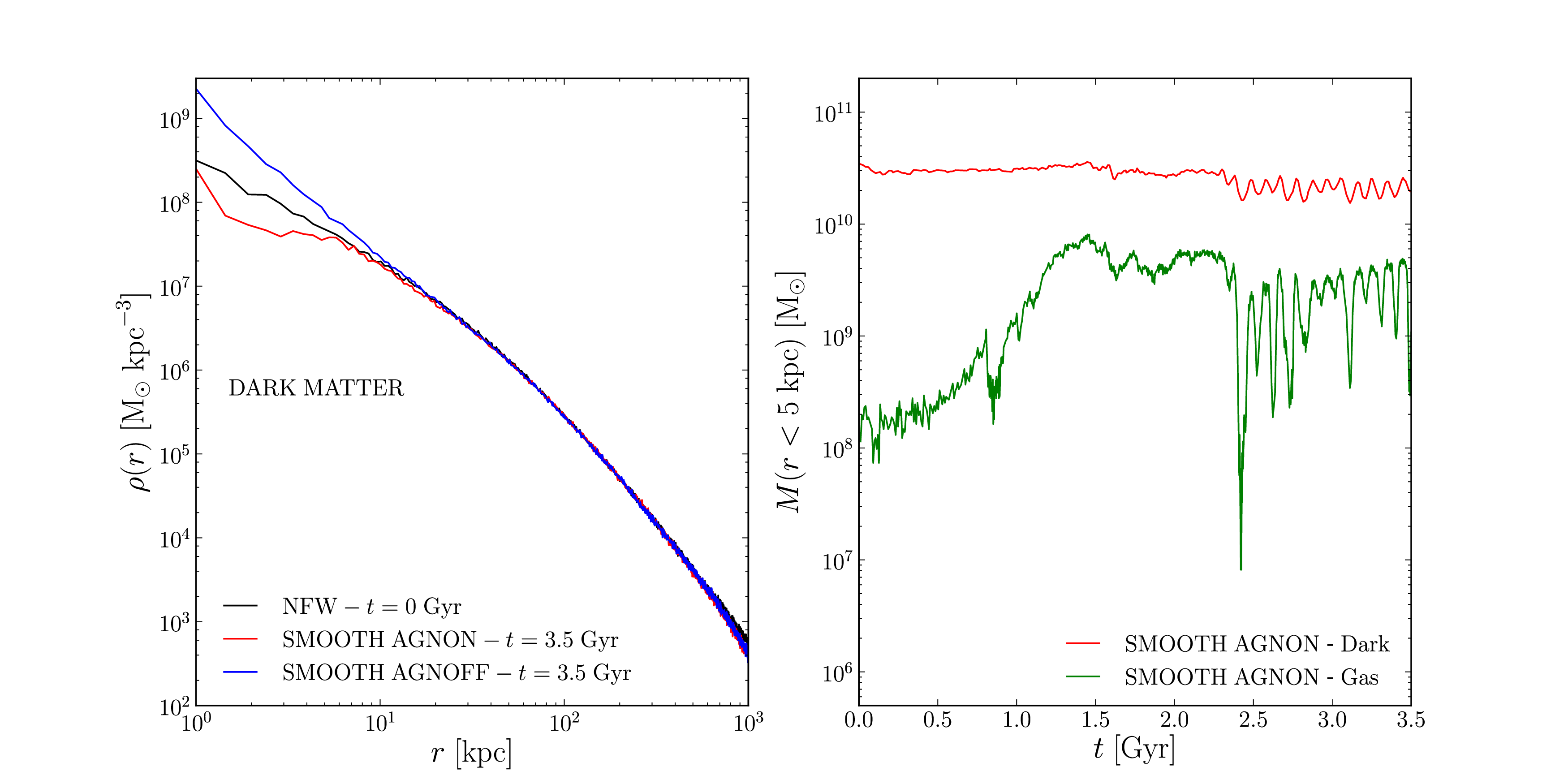}
  \caption{Left: dark matter density profiles at different ages and for the SMOOTH AGNOFF and SMOOTH AGNON simulations which do not include gas clumps. The peak of the dark matter distribution is chosen as a center. In the AGNOFF case we observe the effects of adiabatic contraction. In the AGNON case the effect of the heating-ejection mechanism is very weak compared to the case with clumps.  Right:  evolution of the dark (red line) and gaseous (green line) mass within 5 kpc from the cluster center in the SMOOTH AGNON simulation.}
  \label{fig:smooth}
\end{figure*}

To understand the importance of clumpy accretion, we also carried out simulations using a smooth halo without infalling gas clouds. These simulations showed that it is not possible to trigger heating-ejection cycles when low metallicity ($Z\sim0.01Z_{\odot}$) is adopted: the cooling time is so long that the gas cannot cool efficiently. AGN feedback stays active in its weak radio mode, which provides enough energy to maintain equilibrium and the original gas distribution hardly evolves. The situation is different if a higher metallicity is taken for the gas. 

Figure~\ref{fig:smooth} shows the results we obtain choosing $Z=0.2 Z_{\odot}$ in the SMOOTH AGNON and SMOOTH AGNOFF simulations. For this choice of gas metallicity, the cooling time is comparable to that of the FAST and SLOW simulations after the gas in the clumps mixed with the gaseous halo. In the SMOOTH AGNOFF case gas cools fast enough to produce contraction of the dark matter profile and steepening of the central cusp after 3.5 Gyr. When AGN feedback is turned on, there is a slight flattening of the initial central cusp after 3.5 Gyrs, and within $\sim 3$ kpc from the centre. The right panel of Figure~\ref{fig:smooth} shows how the mass enclosed within 5 kpc evolves in the SMOOTH AGNON run. Gas infall characterizes the first 1.5 Gyr of evolution until AGN activity becomes strong enough to prevent further infall. Since dark matter still dominates the central regions of the halo, no significant variation of the enclosed dark matter mass is observed as a response to gas motions. By $t= 2$ Gyr the black hole is massive enough to trigger some weak heating-ejection cycles. 

These results need to be carefully interpreted. The region affected by the AGN in these "cooling flow" simulations, is close to our resolution limit. Our adopted halo properties are chosen to correspond to a high redshift cluster progenitor, rather than a massive cooling flow cluster. 
The metallicity for the above simulation, that shows a mild evolution, is therefore rather high. 

\section{Summary}
\label{sec:summary}

We analysed the results of N-body hydrodynamical simulations of a dark matter halo of mass $M_{200}=1.42 \times 10^{13}$~M$_{\odot}$ in the presence of gas cooling and AGN feedback. Our initial conditions are chosen to represent the early evolution of the mass distribution in cluster sized halos. Progenitors of halos of mass $\sim 10^{14}$ M$_{\odot}$ identified at redshift $z=0$ have a typical mass of a few $10^{13}$ M$_{\odot}$ at redshift $z\gtrsim 2$. A low metallicity gaseous halo is embedded in the dark matter distributions and we provide additional fuel for the central SMBH by letting ten high metallicity cold gas clumps fall into the central region. In all our runs, we consider radiative gas cooling. In two of our runs we also set a SMBH at the centre of the halo and we implement AGN feedback following a modification of the \cite{Booth:2009p501} already adopted by \cite{2011MNRAS.414..195T}.  

The aim of this work was to further investigate and confirm the results of the cosmological hydrodynamical simulations of \citep{2012MNRAS.422.3081M}, in which the central distribution of stars and dark matter within a cD galaxy were modified through violent gas motions induced by AGN feedback. These results are similar to the those obtained by other authors using supernovae feedback in low mass dark matter halos \citep{Governato:2010p1442, 2012MNRAS.421.3464P, 2011arXiv1111.5620M, 2012arXiv1206.4895T}. 

There are limitations to our study. The resolution of our simulations is approximately 0.5 kpc which could be improved in the future. Furthermore, compared to the observational data, our simulations over-predict the size of the regions in which the shallowing of dark matter mass profiles is measured. The simple AGN feedback model we adopt is probably too efficient in unbinding baryons from the central regions of halos. Given these considerations, our results should be considered as a demonstration of the maximum effect that AGN feedback can have on the mass distribution in halos hosting galaxy clusters. 

Given these caveats, we can summarise our results as follows:

\begin{itemize}

\item In the absence of AGN feedback, the gas clumps fall into the central region of the halo and mix with the rest of the gas. Given that the clumps provide high metallicity material, the result is a decrease of the cooling time in the central regions of the halo. Radiative cooling leads to the condensation of gas at the halo centre. The dark matter halo responds to the gas condensation via adiabatic contraction, increasing the central density. 

\item In the presence of AGN feedback, after 1--2 Gyr we find a reduction
in the dark matter density in the central regions of the halo. Roughly 60~\% of the dark mass within 5 kpc from the center is moved to larger radii. The result is a mass distribution characterized by a dark matter density profile with a density slope in the region $r\lesssim 5-10$~kpc significantly shallower than that of the initial NFW model.

\item The only process responsible for the dark mass depletion in the central regions of the halo is related to AGN feedback and the resulting gas dynamics. After the gas clumps fall into the central region of the halo and mix with halo gas, we observe cycles of fast accretion onto the SMBH, followed by AGN feedback activity that blows gas away. The expelled gas cools and falls back into the central region of the halo. The whole cycle happens on a timescale comparable to the dynamical time. Repeated episodes generate a series of gravitational potential fluctuations that influence the total mass distribution in an irreversible way \citep{2012MNRAS.421.3464P}. 

\item If we use a smooth gas distribution without infalling gas clumps we are not able to trigger the mechanism we just described unless we assume high gas metallicity (shorter cooling time). The observed effect on the dark matter profile is then rather small. High central gas densities that arise from infall emerges as a key element to produce cusp-core transformations. At high redshift this may be triggered through mergers, gas stripping and fueling from cold streams.
\end{itemize}

Recent observational work \citep{2004ApJ...604...88S, 2008ApJ...674..711S, 2009ApJ...706.1078N, 2011ApJ...728L..39N, 2011A&A...531A.119R, 2012arXiv1209.1391N} shows that cluster with shallow central dark matter density profiles may have been observed. Furthermore, very massive elliptical galaxies are observed to have surface brightness profiles with shallow central slopes \citep{1999ASPC..182..124K, 2000ApJS..128...85Q, 2003AJ....125..478L, 2004AJ....127.1917T, 2005AJ....129.2138L, 2007ApJ...671.1456C,  2009ApJS..182..216K, 2011arXiv1108.0997G}. 

The two phenomena might be related, given that dark matter and stars can both be influenced by the same physical processes. Indeed, a strong prediction from our past simulations is that the sizes of the central cores in the dark matter and the stars are similar \citep{2012MNRAS.422.3081M}. Additional observational evidence would be important to clarify this claim and to reveal possible correlations between AGN activity, stellar and dark matter cores and the global properties of clusters and the intra-cluster medium.

\section*{Acknowledgments}
We thank our referee for many suggestions that improved the quality of our paper.
The simulations were performed on the Schr\"{o}dinger cluster at the University of Z\"{u}rich.


\bibliography{papers}

\begin{thebibliography}{}

\bibitem[\protect\citeauthoryear{Booth \& Schaye}{Booth \&
  Schaye}{2009}]{Booth:2009p501}
Booth C.~M.,  Schaye J.,  2009, Monthly Notices of the Royal Astronomical
  Society, 398, 53

\bibitem[\protect\citeauthoryear{{Bournaud}, {Juneau}, {Le Floc'h}, {Mullaney},
  {Daddi}, {Dekel}, {Duc}, {Elbaz}, {Salmi} \& {Dickinson}}{{Bournaud}
  et~al.}{2012}]{2012ApJ...757...81B}
{Bournaud} F.,  {Juneau} S.,  {Le Floc'h} E.,  {Mullaney} J.,  {Daddi} E.,
  {Dekel} A.,  {Duc} P.-A.,  {Elbaz} D.,  {Salmi} F.,    {Dickinson} M.,  2012,
  \apj, 757, 81

\bibitem[\protect\citeauthoryear{{C{\^o}t{\'e}}, {Ferrarese}, {Jord{\'a}n},
  {Blakeslee}, {Chen}, {Infante}, {Merritt}, {Mei}, {Peng}, {Tonry}, {West} \&
  {West}}{{C{\^o}t{\'e}} et~al.}{2007}]{2007ApJ...671.1456C}
{C{\^o}t{\'e}} P.,  {Ferrarese} L.,  {Jord{\'a}n} A.,  {Blakeslee} J.~P.,
  {Chen} C.-W.,  {Infante} L.,  {Merritt} D.,  {Mei} S.,  {Peng} E.~W.,
  {Tonry} J.~L.,  {West} A.~A.,    {West} M.~J.,  2007, \apj, 671, 1456

\bibitem[\protect\citeauthoryear{{Del Popolo}}{{Del
  Popolo}}{2009}]{2009ApJ...698.2093D}
{Del Popolo} A.,  2009, \apj, 698, 2093

\bibitem[\protect\citeauthoryear{{Del Popolo}}{{Del
  Popolo}}{2012}]{2012MNRAS.424...38D}
{Del Popolo} A.,  2012, \mnras, 424, 38

\bibitem[\protect\citeauthoryear{{Dubois}, {Devriendt}, {Slyz} \&
  {Teyssier}}{{Dubois} et~al.}{2012}]{2012MNRAS.420.2662D}
{Dubois} Y.,  {Devriendt} J.,  {Slyz} A.,    {Teyssier} R.,  2012, \mnras, 420,
  2662

\bibitem[\protect\citeauthoryear{{Duffy}, {Schaye}, {Kay}, {Dalla Vecchia},
  {Battye} \& {Booth}}{{Duffy} et~al.}{2010}]{2010MNRAS.405.2161D}
{Duffy} A.~R.,  {Schaye} J.,  {Kay} S.~T.,  {Dalla Vecchia} C.,  {Battye}
  R.~A.,    {Booth} C.~M.,  2010, \mnras, 405, 2161

\bibitem[\protect\citeauthoryear{{El-Zant}, {Hoffman}, {Primack}, {Combes} \&
  {Shlosman}}{{El-Zant} et~al.}{2004}]{2004ApJ...607L..75E}
{El-Zant} A.~A.,  {Hoffman} Y.,  {Primack} J.,  {Combes} F.,    {Shlosman} I.,
  2004, \apjl, 607, L75

\bibitem[\protect\citeauthoryear{Fromang, Hennebelle \& Teyssier}{Fromang
  et~al.}{2006}]{Fromang:2006p400}
Fromang S.,  Hennebelle P.,    Teyssier R.,  2006, Astronomy and Astrophysics,
  457, 371

\bibitem[\protect\citeauthoryear{{Gnedin} \& {Zhao}}{{Gnedin} \&
  {Zhao}}{2002}]{2002MNRAS.333..299G}
{Gnedin} O.~Y.,  {Zhao} H.,  2002, \mnras, 333, 299

\bibitem[\protect\citeauthoryear{{Goerdt}, {Moore}, {Read} \&
  {Stadel}}{{Goerdt} et~al.}{2010}]{2010ApJ...725.1707G}
{Goerdt} T.,  {Moore} B.,  {Read} J.~I.,    {Stadel} J.,  2010, \apj, 725, 1707

\bibitem[\protect\citeauthoryear{Governato, Brook, Mayer, Brooks, Rhee,
  Wadsley, Jonsson, Willman, Stinson, Quinn \& Madau}{Governato
  et~al.}{2010}]{Governato:2010p1442}
Governato F.,  Brook C.,  Mayer L.,  Brooks A.,  Rhee G.,  Wadsley J.,  Jonsson
  P.,  Willman B.,  Stinson G.,  Quinn T.,    Madau P.,  2010, Nature, 463, 203

\bibitem[\protect\citeauthoryear{{Graham}}{{Graham}}{2011}]{2011arXiv1108.0997%
G}
{Graham} A.~W.,  2011, ArXiv e-prints

\bibitem[\protect\citeauthoryear{{Komatsu} \& {Seljak}}{{Komatsu} \&
  {Seljak}}{2001}]{2001MNRAS.327.1353K}
{Komatsu} E.,  {Seljak} U.,  2001, \mnras, 327, 1353

\bibitem[\protect\citeauthoryear{{Kormendy}}{{Kormendy}}{1999}]{1999ASPC..182.%
.124K}
{Kormendy} J.,  1999, in {D.~R.~Merritt, M.~Valluri, \& J.~A.~Sellwood} ed.,
  Galaxy Dynamics - A Rutgers Symposium Vol.~182 of Astronomical Society of the
  Pacific Conference Series, {The Central Structure of Elliptical Galaxies and
  the Stellar-Dynamical Search for Supermassive Black Holes}.
pp 124--+

\bibitem[\protect\citeauthoryear{{Kormendy}, {Fisher}, {Cornell} \&
  {Bender}}{{Kormendy} et~al.}{2009}]{2009ApJS..182..216K}
{Kormendy} J.,  {Fisher} D.~B.,  {Cornell} M.~E.,    {Bender} R.,  2009, \apjs,
  182, 216

\bibitem[\protect\citeauthoryear{Krumholz, McKee \& Klein}{Krumholz
  et~al.}{2004}]{Krumholz:2004p1079}
Krumholz M.~R.,  McKee C.~F.,    Klein R.~I.,  2004, The Astrophysical Journal,
  611, 399

\bibitem[\protect\citeauthoryear{{Laine}, {van der Marel}, {Lauer}, {Postman},
  {O'Dea} \& {Owen}}{{Laine} et~al.}{2003}]{2003AJ....125..478L}
{Laine} S.,  {van der Marel} R.~P.,  {Lauer} T.~R.,  {Postman} M.,  {O'Dea}
  C.~P.,    {Owen} F.~N.,  2003, \aj, 125, 478

\bibitem[\protect\citeauthoryear{{Laporte}, {White}, {Naab}, {Ruszkowski} \&
  {Springel}}{{Laporte} et~al.}{2012}]{2012MNRAS.424..747L}
{Laporte} C.~F.~P.,  {White} S.~D.~M.,  {Naab} T.,  {Ruszkowski} M.,
  {Springel} V.,  2012, \mnras, 424, 747

\bibitem[\protect\citeauthoryear{{Lauer}, {Faber}, {Gebhardt}, {Richstone},
  {Tremaine}, {Ajhar}, {Aller}, {Bender}, {Dressler}, {Filippenko}, {Green},
  {Grillmair}, {Ho}, {Kormendy}, {Magorrian}, {Pinkney} \& {Siopis}}{{Lauer}
  et~al.}{2005}]{2005AJ....129.2138L}
{Lauer} T.~R.,  {Faber} S.~M.,  {Gebhardt} K.,  {Richstone} D.,  {Tremaine} S.,
   {Ajhar} E.~A.,  {Aller} M.~C.,  {Bender} R.,  {Dressler} A.,  {Filippenko}
  A.~V.,  {Green} R.,  {Grillmair} C.~J.,  {Ho} L.~C.,  {Kormendy} J.,
  {Magorrian} J.,  {Pinkney} J.,    {Siopis} C.,  2005, \aj, 129, 2138

\bibitem[\protect\citeauthoryear{{Maccio'}, {Stinson}, {Brook}, {Wadsley},
  {Couchman}, {Shen}, {Gibson} \& {Quinn}}{{Maccio'}
  et~al.}{2011}]{2011arXiv1111.5620M}
{Maccio'} A.~V.,  {Stinson} G.,  {Brook} C.~B.,  {Wadsley} J.,  {Couchman}
  H.~M.~P.,  {Shen} S.,  {Gibson} B.~K.,    {Quinn} T.,  2011, ArXiv e-prints

\bibitem[\protect\citeauthoryear{{Martizzi}, {Teyssier} \& {Moore}}{{Martizzi}
  et~al.}{2012}]{2012MNRAS.420.2859M}
{Martizzi} D.,  {Teyssier} R.,    {Moore} B.,  2012, \mnras, 420, 2859

\bibitem[\protect\citeauthoryear{{Martizzi}, {Teyssier}, {Moore} \&
  {Wentz}}{{Martizzi} et~al.}{2012}]{2012MNRAS.422.3081M}
{Martizzi} D.,  {Teyssier} R.,  {Moore} B.,    {Wentz} T.,  2012, \mnras, 422,
  3081

\bibitem[\protect\citeauthoryear{{Mashchenko}, {Couchman} \&
  {Wadsley}}{{Mashchenko} et~al.}{2006}]{2006Natur.442..539M}
{Mashchenko} S.,  {Couchman} H.~M.~P.,    {Wadsley} J.,  2006, \nat, 442, 539

\bibitem[\protect\citeauthoryear{{Mashchenko}, {Wadsley} \&
  {Couchman}}{{Mashchenko} et~al.}{2008}]{2008Sci...319..174M}
{Mashchenko} S.,  {Wadsley} J.,    {Couchman} H.~M.~P.,  2008, Science, 319,
  174

\bibitem[\protect\citeauthoryear{{Navarro}, {Eke} \& {Frenk}}{{Navarro}
  et~al.}{1996}]{1996MNRAS.283L..72N}
{Navarro} J.~F.,  {Eke} V.~R.,    {Frenk} C.~S.,  1996, \mnras, 283, L72

\bibitem[\protect\citeauthoryear{{Newman}, {Treu}, {Ellis} \& {Sand}}{{Newman}
  et~al.}{2011}]{2011ApJ...728L..39N}
{Newman} A.~B.,  {Treu} T.,  {Ellis} R.~S.,    {Sand} D.~J.,  2011, \apjl, 728,
  L39+

\bibitem[\protect\citeauthoryear{{Newman}, {Treu}, {Ellis}, {Sand}, {Nipoti},
  {Richard} \& {Jullo}}{{Newman} et~al.}{2012}]{2012arXiv1209.1391N}
{Newman} A.~B.,  {Treu} T.,  {Ellis} R.~S.,  {Sand} D.~J.,  {Nipoti} C.,
  {Richard} J.,    {Jullo} E.,  2012, ArXiv e-prints

\bibitem[\protect\citeauthoryear{{Newman}, {Treu}, {Ellis}, {Sand}, {Richard},
  {Marshall}, {Capak} \& {Miyazaki}}{{Newman}
  et~al.}{2009}]{2009ApJ...706.1078N}
{Newman} A.~B.,  {Treu} T.,  {Ellis} R.~S.,  {Sand} D.~J.,  {Richard} J.,
  {Marshall} P.~J.,  {Capak} P.,    {Miyazaki} S.,  2009, \apj, 706, 1078

\bibitem[\protect\citeauthoryear{{Pe{\~n}arrubia}, {Pontzen}, {Walker} \&
  {Koposov}}{{Pe{\~n}arrubia} et~al.}{2012}]{2012ApJ...759L..42P}
{Pe{\~n}arrubia} J.,  {Pontzen} A.,  {Walker} M.~G.,    {Koposov} S.~E.,  2012,
  \apjl, 759, L42

\bibitem[\protect\citeauthoryear{{Peirani}, {Kay} \& {Silk}}{{Peirani}
  et~al.}{2008}]{2008A&A...479..123P}
{Peirani} S.,  {Kay} S.,    {Silk} J.,  2008, \aap, 479, 123

\bibitem[\protect\citeauthoryear{{Pontzen} \& {Governato}}{{Pontzen} \&
  {Governato}}{2012}]{2012MNRAS.421.3464P}
{Pontzen} A.,  {Governato} F.,  2012, \mnras, 421, 3464

\bibitem[\protect\citeauthoryear{{Quillen}, {Bower} \& {Stritzinger}}{{Quillen}
  et~al.}{2000}]{2000ApJS..128...85Q}
{Quillen} A.~C.,  {Bower} G.~A.,    {Stritzinger} M.,  2000, \apjs, 128, 85

\bibitem[\protect\citeauthoryear{{Ragone-Figueroa}, {Granato} \&
  {Abadi}}{{Ragone-Figueroa} et~al.}{2012}]{2012arXiv1202.1527R}
{Ragone-Figueroa} C.,  {Granato} G.~L.,    {Abadi} M.~G.,  2012, ArXiv e-prints

\bibitem[\protect\citeauthoryear{{Read} \& {Gilmore}}{{Read} \&
  {Gilmore}}{2005}]{2005MNRAS.356..107R}
{Read} J.~I.,  {Gilmore} G.,  2005, \mnras, 356, 107

\bibitem[\protect\citeauthoryear{{Richtler}, {Salinas}, {Misgeld}, {Hilker},
  {Hau}, {Romanowsky}, {Schuberth} \& {Spolaor}}{{Richtler}
  et~al.}{2011}]{2011A&A...531A.119R}
{Richtler} T.,  {Salinas} R.,  {Misgeld} I.,  {Hilker} M.,  {Hau} G.~K.~T.,
  {Romanowsky} A.~J.,  {Schuberth} Y.,    {Spolaor} M.,  2011, \aap, 531, A119+

\bibitem[\protect\citeauthoryear{{Rocha}, {Peter}, {Bullock}, {Kaplinghat},
  {Garrison-Kimmel}, {Onorbe} \& {Moustakas}}{{Rocha}
  et~al.}{2012}]{2012arXiv1208.3025R}
{Rocha} M.,  {Peter} A.~H.~G.,  {Bullock} J.~S.,  {Kaplinghat} M.,
  {Garrison-Kimmel} S.,  {Onorbe} J.,    {Moustakas} L.~A.,  2012, ArXiv
  e-prints

\bibitem[\protect\citeauthoryear{{Sand}, {Treu}, {Ellis}, {Smith} \&
  {Kneib}}{{Sand} et~al.}{2008}]{2008ApJ...674..711S}
{Sand} D.~J.,  {Treu} T.,  {Ellis} R.~S.,  {Smith} G.~P.,    {Kneib} J.-P.,
  2008, \apj, 674, 711

\bibitem[\protect\citeauthoryear{{Sand}, {Treu}, {Smith} \& {Ellis}}{{Sand}
  et~al.}{2004}]{2004ApJ...604...88S}
{Sand} D.~J.,  {Treu} T.,  {Smith} G.~P.,    {Ellis} R.~S.,  2004, \apj, 604,
  88

\bibitem[\protect\citeauthoryear{Sutherland \& Dopita}{Sutherland \&
  Dopita}{1993}]{sutherland_dopita93}
Sutherland R.~S.,  Dopita M.~A.,  1993, \apjs, 88, 253

\bibitem[\protect\citeauthoryear{Teyssier}{Teyssier}{2002}]{Teyssier:2002p451}
Teyssier R.,  2002, Astronomy and Astrophysics, 385, 337

\bibitem[\protect\citeauthoryear{Teyssier, Fromang \& Dormy}{Teyssier
  et~al.}{2006}]{Teyssier:2006p413}
Teyssier R.,  Fromang S.,    Dormy E.,  2006, Journal of Computational Physics,
  218, 44

\bibitem[\protect\citeauthoryear{{Teyssier}, {Moore}, {Martizzi}, {Dubois} \&
  {Mayer}}{{Teyssier} et~al.}{2011}]{2011MNRAS.414..195T}
{Teyssier} R.,  {Moore} B.,  {Martizzi} D.,  {Dubois} Y.,    {Mayer} L.,  2011,
  \mnras, 414, 195

\bibitem[\protect\citeauthoryear{{Teyssier}, {Pontzen}, {Dubois} \&
  {Read}}{{Teyssier} et~al.}{2012}]{2012arXiv1206.4895T}
{Teyssier} R.,  {Pontzen} A.,  {Dubois} Y.,    {Read} J.,  2012, ArXiv e-prints

\bibitem[\protect\citeauthoryear{Toro, Spruce \& Speares}{Toro
  et~al.}{1994}]{Toro:1994p1151}
Toro E.~F.,  Spruce M.,    Speares W.,  1994, Shock Waves, 4, 25

\bibitem[\protect\citeauthoryear{Tremaine, Gebhardt, Bender, Bower, Dressler,
  Faber, Filippenko, Green, Grillmair, Ho, Kormendy, Lauer, Magorrian, Pinkney
  \& Richstone}{Tremaine et~al.}{2002}]{tremaine_etal04}
Tremaine S.,  Gebhardt K.,  Bender R.,  Bower G.,  Dressler A.,  Faber S.~M.,
  Filippenko A.~V.,  Green R.,  Grillmair C.,  Ho L.~C.,  Kormendy J.,  Lauer
  T.~R.,  Magorrian J.,  Pinkney J.,    Richstone D.,  2002, \apj, 574, 740

\bibitem[\protect\citeauthoryear{{Trujillo}, {Erwin}, {Asensio Ramos} \&
  {Graham}}{{Trujillo} et~al.}{2004}]{2004AJ....127.1917T}
{Trujillo} I.,  {Erwin} P.,  {Asensio Ramos} A.,    {Graham} A.~W.,  2004, \aj,
  127, 1917

\bibitem[\protect\citeauthoryear{{Vogelsberger}, {Zavala} \&
  {Loeb}}{{Vogelsberger} et~al.}{2012}]{2012MNRAS.423.3740V}
{Vogelsberger} M.,  {Zavala} J.,    {Loeb} A.,  2012, \mnras, 423, 3740

\end{thebibliography}


\label{lastpage}
\end{document}